\documentclass[conference]{IEEEtran}
\IEEEoverridecommandlockouts
\usepackage{cite}
\usepackage{amsmath,amssymb,amsfonts}
\usepackage{algorithmic}
\usepackage{graphicx}
\usepackage{textcomp}
\usepackage{xcolor}
\usepackage{subfigure}
\usepackage{multirow}
\usepackage{color,soul} 
\def\BibTeX{{\rm B\kern-.05em{\sc i\kern-.025em b}\kern-.08em
    T\kern-.1667em\lower.7ex\hbox{E}\kern-.125emX}}
\begin{document}

\title{Millimeter-Wave Path Loss at 73 GHz in Indoor and Outdoor Airport Environments\\
}
\author{\IEEEauthorblockN{Mahfuza Khatun$^{\dag}$, Changyu Guo$^{\dag}$, Letizia Moro$^{\dag}$, David Matolak$^{\ddag}$, Hani Mehrpouyan$^{\dag}$}\\
	
	\IEEEauthorblockA{$^{\dag}$Department of Electrical and Computer Engineering, Boise State University \\
		$^\ddag$Department of Electrical Engineering, University of South Carolina  \\
		Email: \{mahfuzakhatun, changyuguo, letiziamoro, hanimehrpouyan\}@boisestate.edu, matolak@cec.sc.edu	  
	}
}

\maketitle

\begin{abstract}
In this paper, two large-scale fading path loss models are presented based on indoor and outdoor channel measurements at $73$ GHz. The line-of-sight millimeter-wave propagation measurement campaigns were \textit{uniquely} conducted within the indoor and outdoor environments at an airport setting, i.e., the Boise Airport. The channel measurements were made with directional transmit and receive antennas with a $24$ dBi gain at different receive antenna heights. From the measured data, we obtained the parameters of two path loss models, i.e., the close-in reference distance model (CIM) and the floating-intercept model (FIM). Results show that the path loss exponents estimated from the CIM are very close to that of the free-space path loss model, while the FIM provides a better fit to the measurement data.

\end{abstract}


\section{Introduction}

The immense amount of available bandwidth at the millimeter-wave (mmWave) band is an outstanding resource for supporting Gigabit-per-second (Gbps) data rates for backhaul and fronthaul applications in wireless networks. More importantly, the mmWave spectrum can alleviate the spectrum shortage at sub-$6$ GHz frequencies~\cite{rqppaport2013Itwillwork}~\cite{rappaport2015wideband}. 

\subsection{Motivation}
In this paper, we focus our attention on developing channel models for spectrum in the $73$ GHz band. Our research mission is motivated by the shortage of channel measurement data in this band for airports and other environments. It is important to develop such channel models, since the available spectrum in the $73$ GHz band can support more than ten times the data rates of that of the sub-$6$ GHz spectrum. Moreover, to overcome the significant path loss in the $73$ GHz band, an accurate understanding of the channel models that govern this spectrum are needed to devise more appropriate beamforming and antenna structure~\cite{shad2019}~\cite{ almasi2019lens} for use in this band~\cite{rangan2014millimeter}. 

The second focus of this paper is directed at characterizing the $73$ GHz channel models for airport indoor and outdoor settings. This research is motivated by the fact that in the near future, unmanned aerial vehicles (UAVs) will operate in conjunction with piloted airplanes in airport airspaces. More importantly, it is anticipated that automation through the use of machines and robots will affect and improve airport operations in every aspect. To make all of this possible, effective and reliable communication is a necessity. This, in turn will require the use of mmWave bands such as the $73$ GHz spectrum.

In this paper, for the first time, consider channel models at $73$ GHz in indoor and outdoor environments at an airport. Moreover, the resulting measurement data will go a long way in provide the communication research community with a richer data set on channel models in this band. This is important, since different channel measurements obtained by different teams in the mmWave band have resulted in a variety of path loss models for similar propagation environment~\cite{rqppaport2013Itwillwork, samimi20153}. This motivates the need for more channel measurement campaigns in the mmWave band, in this case the $73$ GHz band.  

\subsection{Related Work}
Several companies and research groups have been carrying out channel measurements at mmWave frequencies~\cite{rqppaport2013Itwillwork, rappaport2015wideband, samimi20153}. However, as stated above, there is a lack of actual measurement results for the $73$ GHz frequency band. For instance, the authors in \cite{samimi20153} developed a $3$-D statistical model at both the $28$ GHz and $73$ GHz bands for outdoor line-of-sight (LOS) and non-line of sight (NLOS) environments. Subsequently, we computed the channel parameters at $28$ GHz and $73$ GHz for both LOS and NLOS scenarios in Boise~\cite{mahfuza2017} using the MATLAB statistical channel simulator~\cite{nyusim}. Other than the simulation works mentioned above, in ~\cite{mahfuza2018}, a channel measurement campaigns was conducted to obtain path loss models for $60$ GHz frequency band within airport environments. 

\subsection{Contributions}
To the best of the authors' knowledge, this paper, for the first time, presents an extensive channel measurement campaign at $73$ GHz in an airport environment. The channel measurement campaigns were conducted both outside on the taxiways and the airport tarmac and inside the concourse and gate areas. To collect a set of comparative measurement data, we conducted a second set of measurement campaigns at Boise State University in both indoor and outdoor scenarios. The channel measurements were made with directional transmit and receive antennas with a $24$ dBi gain at different receive antenna heights. From the measured data, we obtained the parameters of two path loss models, i.e., the close-in reference distance model (CIM) and the floating-intercept model (FIM). Results show that the path loss exponents estimated from the CIM are very close to that of the free-space path loss model, while the FIM provides a better fit to the measurement data.

This paper is organized as follows: Section~\ref{sec:measurement_hardware} describes the channel measurement hardware setup, and measurement environments. Section ~\ref{sec:large_scale_fading} presents the large-scale fading channel models used in this paper. In Section ~\ref{sec:results}, we present and analyze the collected measurement data, while, Section ~\ref{sec:conclusion} concludes the paper.
\begin{figure}[t]
	\begin{center} 		
		\includegraphics[width=0.9\linewidth,height=5.5cm]{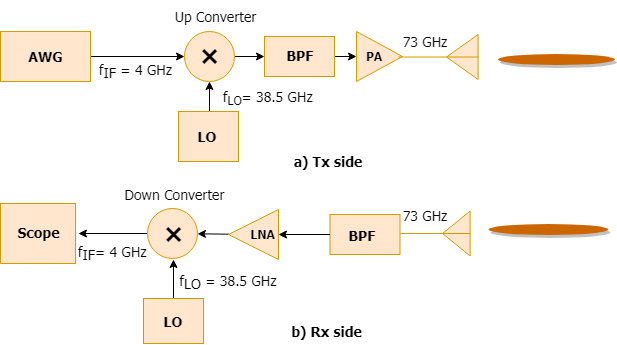}
	\end{center}
	\caption{Block diagram of the (a) Tx and (b) Rx for the mm-Wave propagation measurements at 73 GHz}
	\label{Fig1}
\end{figure}

\section{measurement hardware and procedure}
\label{sec:measurement_hardware}
In this section the measurement hardware and the overall setup for both the indoor and outdoor channel measurement campaigns are described.
\subsection{Measurement Setup}
 The channel measurement hardware setup is shown in Fig.~\ref{Fig1}. At the transmitter, the signal is digitally modulated at $4$ GHz inside the arbitrary waveform generator (AWG), with binary phase shift keying (BPSK) modulation with a $1$ GHz symbol rate. Following this, the RF output from the AWG is connected to intermediate frequency (IF) port of the up-converter mixer. This signal is then further up-converted to $73$ GHz. The mixer uses a local oscillator (LO) that operates at $38.5$ GHz. Following upconversion, a bandpass filter (BPF) is used to suppress the undesired out of band signal components. Finally, a power amplifier (PA) with a gain of $20$ dB is placed before the transmitter antenna. A directional horn antenna with a gain of $24$ dBi and a beamwidth of $7^{\circ}$ elevation and $11^{\circ}$ azimuth is used at both the transmitter and receiver. At the receiver, the antenna is connected a bandpass filter and then a low noise amplifier. A down-converter is used to shift the $73$ GHz signal to the baseband frequency of $4$ GHz. The received signal is then fed to the o-scope, where the received signal strength is estimated. All the hardware specifications are listed in Table~\ref{Tab_1}.

\begin{table}[t]
	\caption{\bf{ Hardware Specification for the $73$ GHz Channel Measurement Campaign}}
	\label{Tab_1}	
	\centering
	\begin{tabular}{|c|c|}
		\hline
		\multicolumn{2}{|c|}{73 GHz Channel Measurement Campaign} \\ \hline
		Carrier Frequency                    & 73 GHz             \\ \hline
		Tx antenna gain                      & 24 dBi             \\ \hline
		Rx antenna gain                      & 24 dBi             \\ \hline
		3dB beamwidth in V-plane             & 7 $^{\circ}$                 \\ \hline
		3dB beamwidth in H-plane             & 11 $^{\circ}$                \\ \hline
		Modulation scheme                    & BPSK               \\ \hline
		Bandwidth                            & 1.3 GHz            \\ \hline
		Max Tx Power                         & 3 dBm              \\ \hline
		Max meas. path loss                 & 112 dB             \\ \hline
	\end{tabular}
\end{table}


\begin{figure}[t]
	\begin{center}
		\subfigure []
		{ \includegraphics[width=0.47\linewidth,height=5cm]{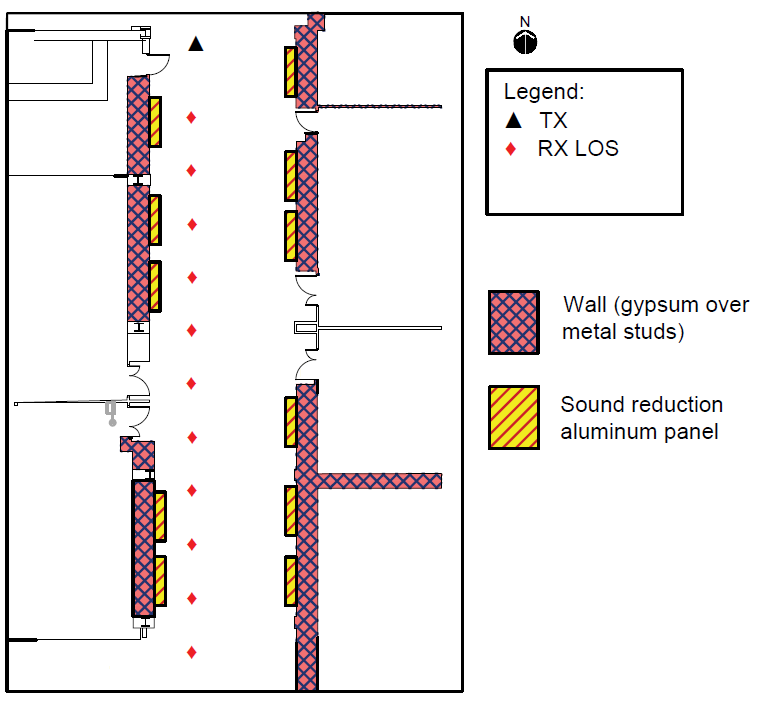}\label{fig.hallway}}
		\subfigure []
		{\includegraphics[width=0.47\linewidth,height=5cm]{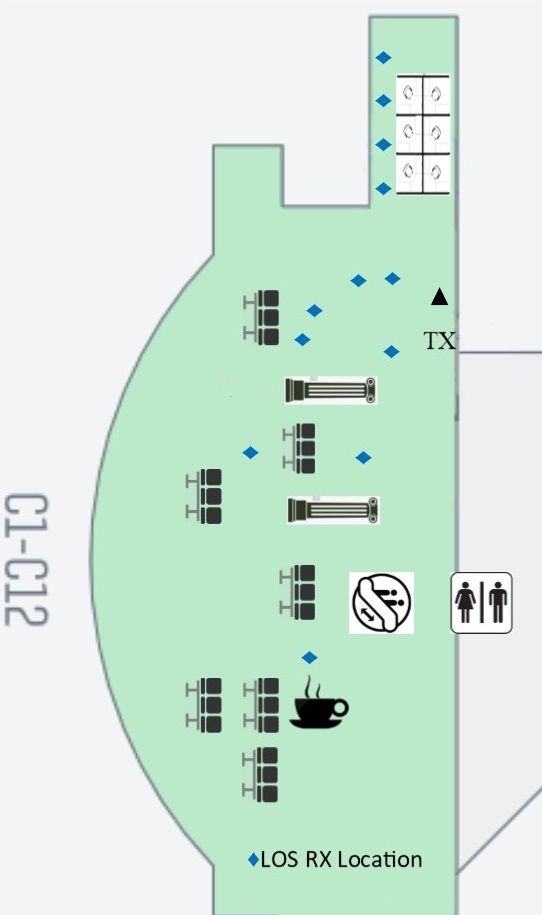}\label{fig.gate}}
	\end{center}
	\caption{(a) Floor map of the Micron Engineering building. (b) The airport concourse C gate.}
	\label{Fig_2}
	
\end{figure}

\subsection{Indoor Measurements}
During September of 2018, our first measurement campaign at $73$ GHz was conducted in the hallway of the Micron Engineering building (MEC) at Boise State University. The overall hallway layout is shown in Fig~\ref{fig.hallway}. The size of the MEC building hallway is about $ 32\times 2.2 \times 1.9$ $m^3$. The walls are made of sheetrock over metal studs, the ceiling tiles are made of a fiberboard material, and the ground is comprised of concrete. The transmitter and receiver were organized with two movable carts equipped with the instruments. The measurements were taken in thirteen different receiver locations while keeping the position of the transmit antenna fixed. The antennas were manually rotated to find the strongest received power for each unique Tx-Rx location. 

The second set of measurements was completed in the indoor gate area of the Boise Airport. The gated area was specifically selected since it is a crowded environment with metallic objects and chairs. The overall layout of the airport gated area has been shown in Fig.~\ref{fig.gate}. The data were collected at three different receiver antenna heights. The transmitter antenna height was fixed at 1.6 meters relative to the ground. 

\begin{figure}[t]
	\begin{center}
		\subfigure [ ]
		{ \includegraphics[width=0.47\linewidth,height=5cm]{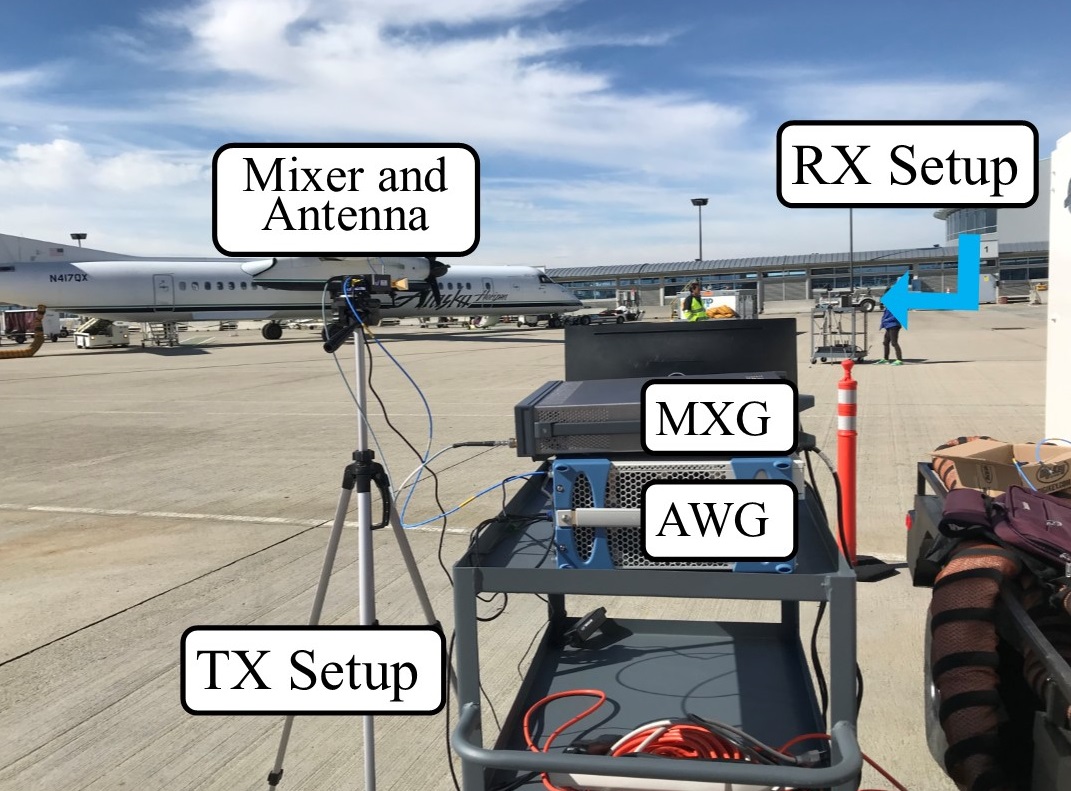}\label{fig.airport_picture}}
		\subfigure []
		{\includegraphics[width=0.47\linewidth,height=5cm]{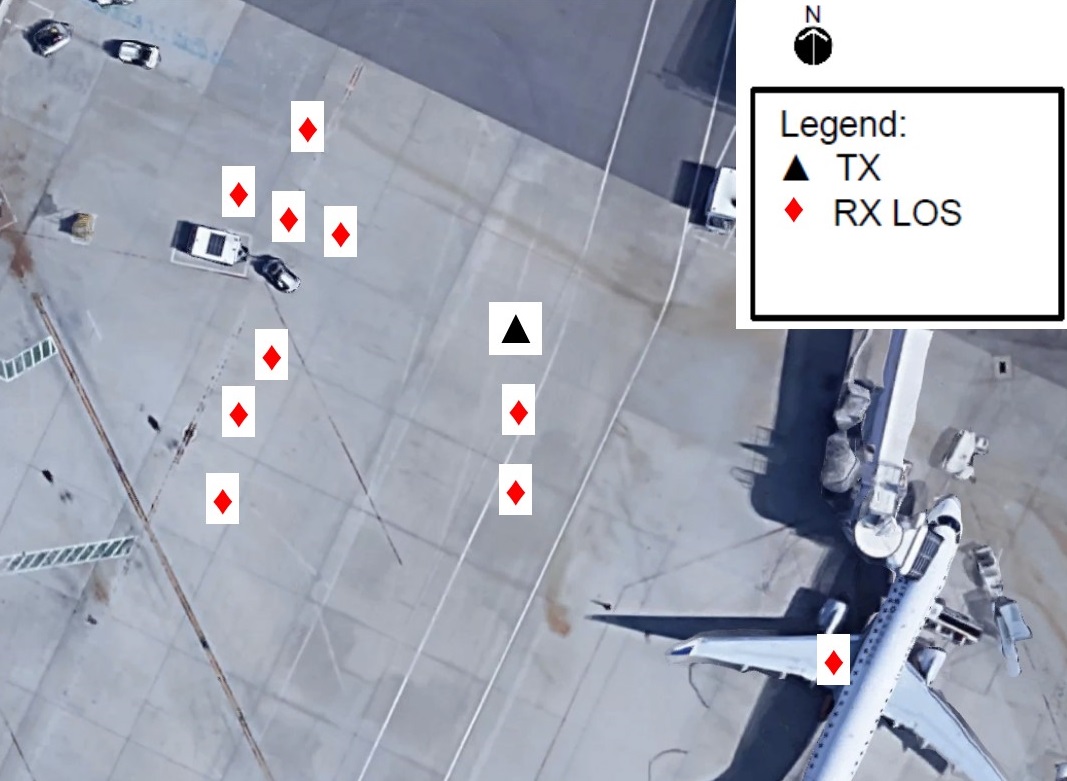}\label{fig.airport_map}}
	\end{center}
	\caption{(a) Photo of the outdoor setting in the airport. (b) Overhead image of the outdoor setting at the Boise Airport showing the transmitter and receiver at various locations.}
	\label{Fig_3}
	
\end{figure}

\subsection{Outdoor Measurements}
In 2018 of September, two outdoor propagation measurement campaigns were conducted at $73$ GHz at the Boise Airport and at the Boise State University. The data capturing methodology is similar to that of the prior subsection. Fig.~\ref{fig.airport_picture} shows the outdoor setting at the airport just outside of the gate areas. The height of the transmitter antenna was fixed at $1.6$ m from the ground level, and ten receiver locations with three antenna heights, i.e., $1.6$ m, $1.4$ m and $1.3$ m, were selected. The airplanes shown in Fig.~\ref{fig.airport_map} was not present during the data collection.

\begin{figure}[t]
	\begin{center}
		\subfigure [ ]
		{ \includegraphics[width=0.47\linewidth,height=5cm]{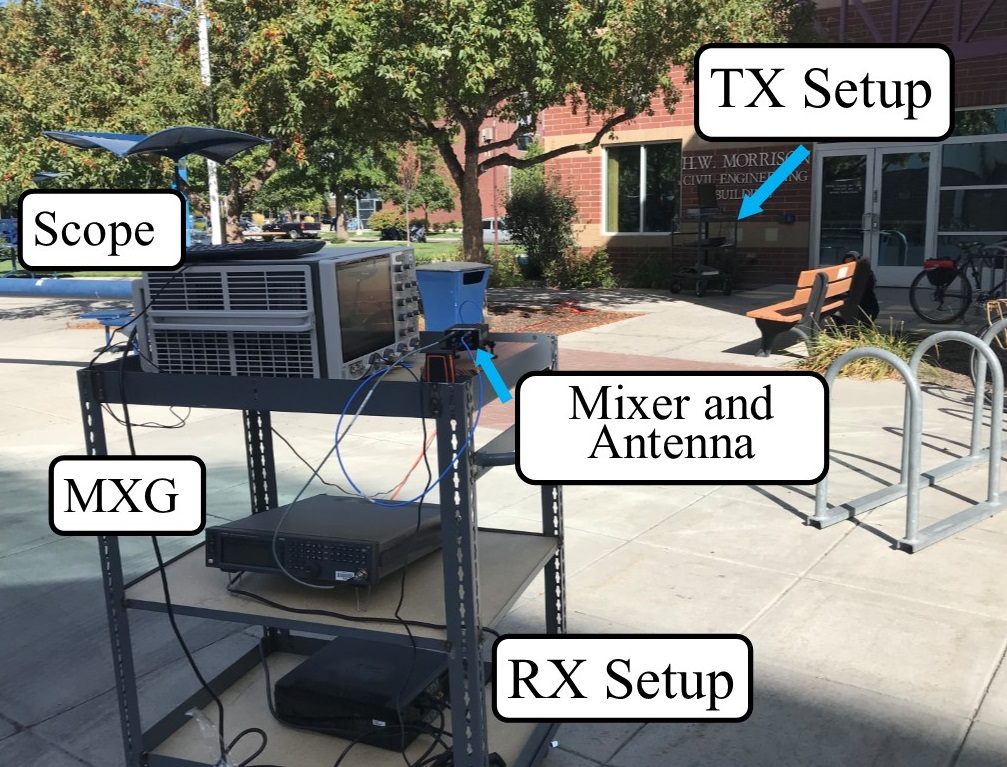}\label{fig.campus_picture}}
		\subfigure []
		{\includegraphics[width=0.47\linewidth,height=5cm]{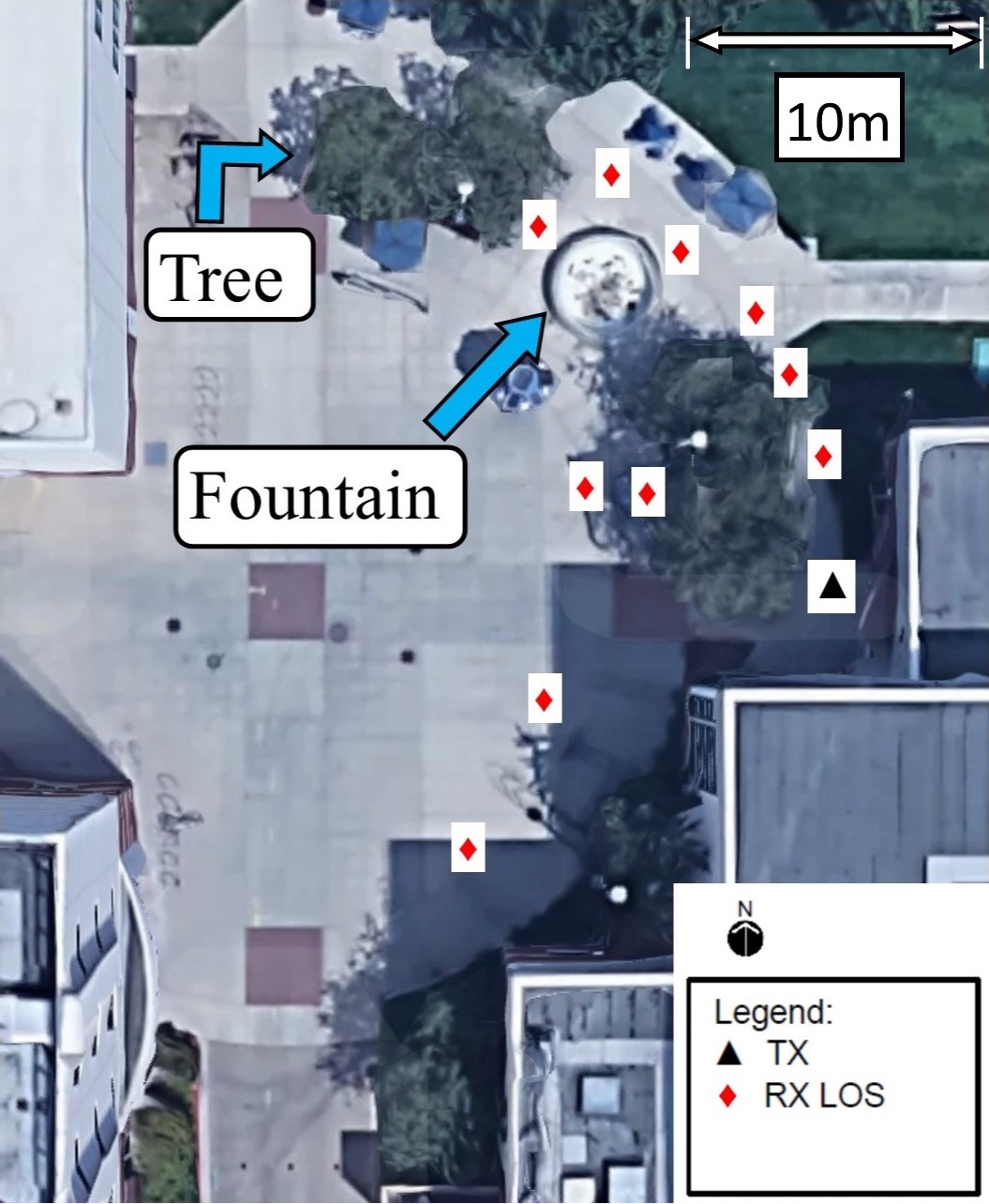}\label{fig.campus_map}}
	\end{center}
	\caption{(a) Photo of the outdoor environment at the Boise State campus. (b) Overhead image of the outdoor environment at Boise State.}
	\label{Fig_4}
	
\end{figure}

In addition, another outdoor campaign was organized at Boise State University for LOS scenarios. The transmitter location was fixed at a height of $1.21 m$ from the ground level and various receiver locations (height of $1.18m$) were selected. The overall layout for this scenario is depicted in Fig.~\ref{Fig_4}. At each transceiver separation , the transmit antenna was manually tilted down toward the receiver antenna, and the receiver antenna was adjusted in such a way as to receive the highest signal to noise ratio. All antennas were placed in vertical-to-vertical (V-V) polarization in both indoor and outdoor scenarios.

\begin{table*}[ht!]
	\renewcommand{\arraystretch}{1.2}
	
	\caption{\bfseries Comparison of key path loss parameters in CIM (1-meter reference distance) and that of FIM for the measurements at Boise Airport and at Boise State University}
	
	\label{tab2}
	\centering
	
	\begin{tabular}{|c|l|l|l|l|l|l|l|l|}
		
		\hline
		\multicolumn{9}{|c|}{Directional Path Loss Models} 
		                                                                                      \\ \hline
		                                                                                     \multirow{2}{*}{Environments}   & \multirow{2}{*}{Scenario} & \multirow{2}{*}{$h_{Tx}$, m} & \multirow{2}{*}{$h_{Rx}$, m }& \multicolumn{2}{c|}{CIM}                     & \multicolumn{3}{c|}{FIM}                             \\ \cline{5-9} 
		                                                                                     &                        &                     &                     & n & $\sigma$, dB& $\alpha$, dB & $ \beta$ & $\sigma$, dB\\ \hline
		                                                                                     Indoor (airport gate) & LOS                    & 1.21                & 1.18                & 2.1                  & 2.6                   & 75    & 1.8                  & 2.4                   \\ \hline
		                                                                                     Indoor (MEC building) & LOS                    & 1.21                & 1.18                & 1.8                  & 2.05                  & 73    & 1.42                 & 1.2                   \\ \hline
		                                                                                     outdoor (MEC building) & LOS                    & 1.2                 & 1.8                 & 1.8                  & 2.2                   & 73    & 1.5                  & 1.8                   \\ \hline
	                                                                                     \end{tabular}
                                                                                      \end{table*}

\begin{table*}[ht!]
	\renewcommand{\arraystretch}{1.5}
	
	\caption{\bfseries The FIM parameters for Boise Airport in both indoor and outdoor scenarios at $73$ GHz for different receiver antenna heights. Transceiver separation ranges from $1$ m to $30$ m}
	\label{tab3}
	\centering
	\begin{tabular}{|c|c|c|c|c|c|c|}
		\hline
		\multirow{2}{*}{Environments} & \multirow{2}{*}{$h_{Tx}$, m} & \multirow{2}{*}{$h_{Rx}$, m} & \multirow{2}{*}{Path Loss Scenarios} & \multicolumn{3}{c|}{Parameters for directional floating-intercept model} \\ \cline{5-7} 
		&                     &                     &                     & $\alpha$, dB    & $\beta$   & $\sigma$, dB   \\ \hline
		
		\multirow{3}{*}{Indoor Airport} & \multirow{3}{*}{1.6} & 1.6      & \multirow{3}{*}{LOS} & 72    & 1.95     & 1.6      \\ \cline{3-3} \cline{5-7} 
		&                & 1.47                 &                      &    77 & 1.6     & 2.4       \\ \cline{3-3} \cline{5-7} 
		&                      & 1.3                &                      & 92   & 0.4      & 6.5      \\ \hline

		\multirow{3}{*}{Outdoor Airport} & \multirow{3}{*}{1.6} & 1.6      & \multirow{3}{*}{LOS} & 73    & 1.7    & 1.24      \\ \cline{3-3} \cline{5-7} 
		&                & 1.47                 &                      &    76 & 1.6     & 2.8       \\ \cline{3-3} \cline{5-7} 
		&                      & 1.3                &                      & 89   & 0.62     & 6.11     \\ \hline
	\end{tabular}
\end{table*}

\begin{figure}[t]
	\begin{center}
		
		\includegraphics[width=0.9\linewidth,keepaspectratio]{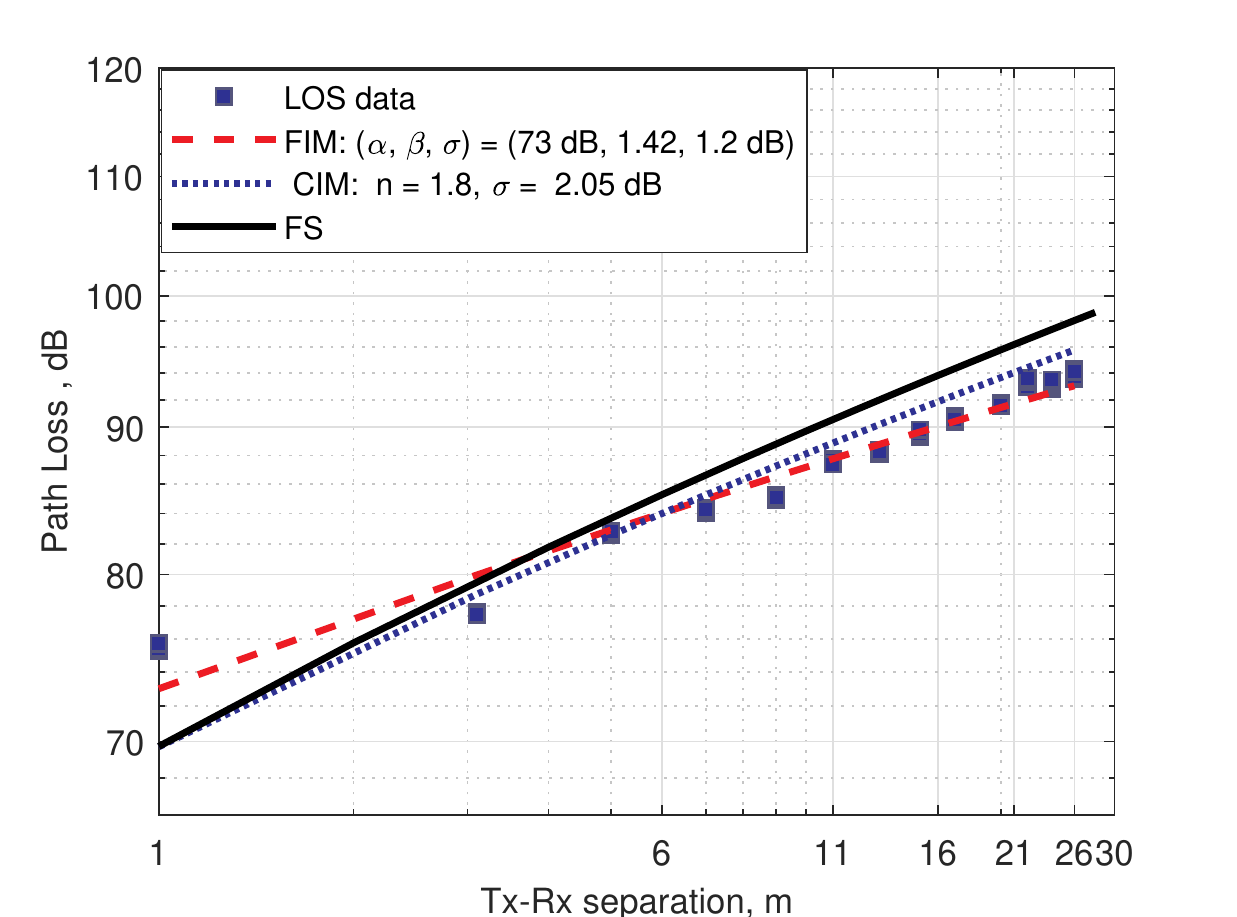}
	\end{center}
	\caption{FIM and CIM along with the measurement data taken from the MEC building in LOS indoor scenario ($h_{tx} = 1.21$ m and  $h_{rx}= 1.18$ m ).  The blue square and solid black line represent the measurement LOS data and free-space path loss at $73$ GHz respectively. The dashed red line and dotted blue line show the FIM and CIM path loss models respectively}
	\label{Fig5}
\end{figure}


\begin{figure}[t]
	\begin{center}
		
		\includegraphics[width=0.9\linewidth,keepaspectratio]{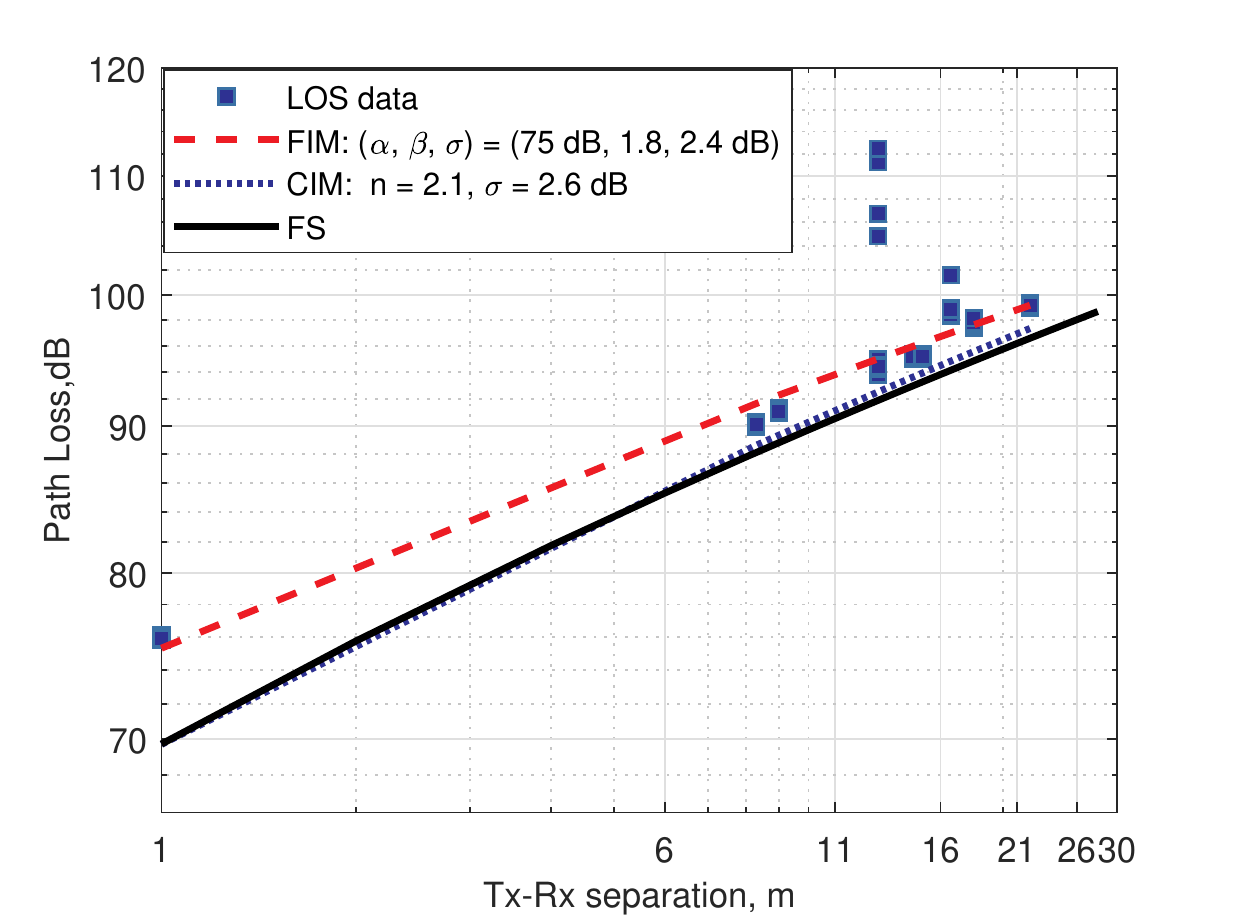}
	\end{center}
	\caption{FIM and CIM along with the measurement data taken from the airport gate in LOS indoor scenario ($h_{tx} = 1.21$ m and  $h_{rx} = 1.18$ m ). The blue square and solid black line represent the measurement LOS data and free-space path loss at 73 GHz respectively. The dashed red line and dotted blue show the FIM and CIM path loss models respectively }
	\label{Fig6}
\end{figure}

\begin{figure}[t]
	\begin{center}
		
		\includegraphics[width=0.9\linewidth,keepaspectratio]{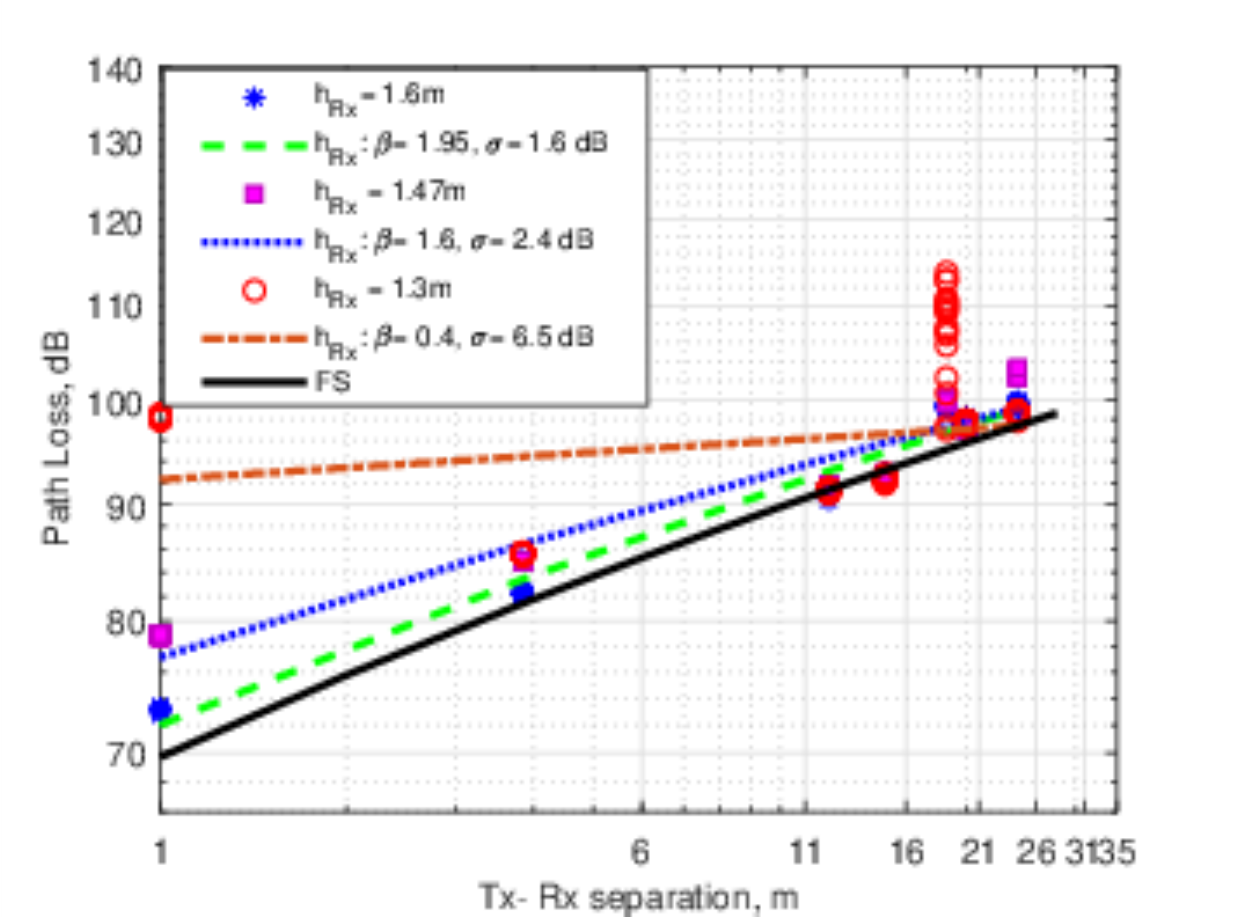}
	\end{center}
	\caption{The FIM model at 73 GHz with different Rx antenna heights from the ground level when the Tx antenna height is kept at $1.6$ m. The blue star, purple square, and red circles represent the measurement LOS data for $ h_{Rx} = 1.6$, $1.47$ and $1.3$ m respectively and the dashed green, dotted blue, and dash dot orange lines show the FIM model corresponding those measurement data collected from the airport gate in LOS indoor scenario. The solid line represents the free-space path loss model at 73 GHz  }
	\label{Fig7}
\end{figure}



\begin{figure}[t]
	\begin{center}
		
		\includegraphics[width=0.9\linewidth,keepaspectratio]{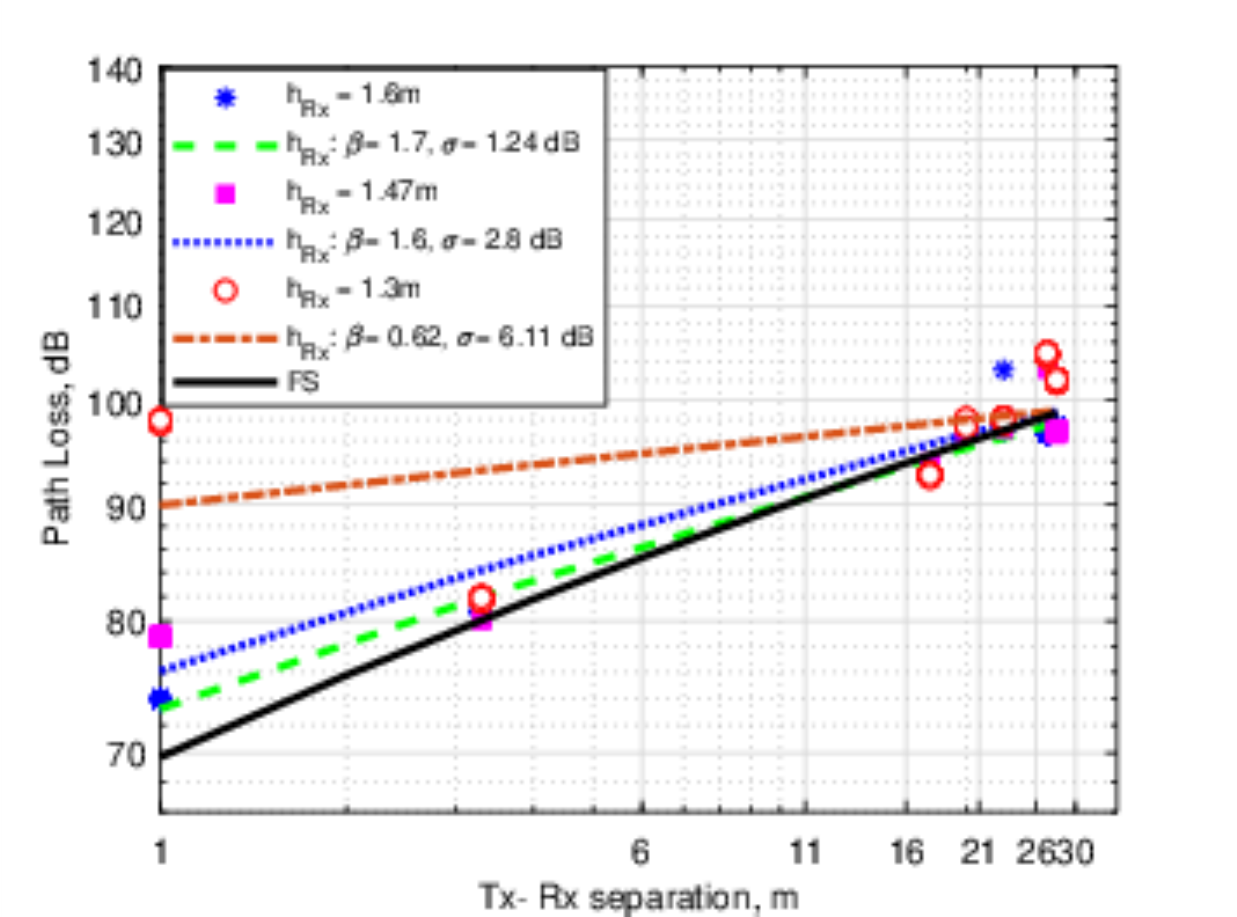}
	\end{center}
	\caption{The FIM model at 73 GHz with different Rx antenna heights from the ground level when the Tx antenna height is kept at $1.6$ m from the ground level. The blue star, purple square, and red circles represent the measurement LOS data for $ h_{Rx} = 1.6$, $1.47$ and $1.3$ m respectively and the dashed green, dotted blue, and dash dot orange lines show the FIM model corresponding those measurement data collected from the airport surface area in LOS outdoor scenario. The solid line represents the free-space path loss model at 73 GHz}
	\label{Fig8}
\end{figure}

\section{Large scale fading models}
\label{sec:large_scale_fading}
The directional path loss channel models are obtained for the indoor and outdoor scenarios using CIM and FIM methods. 

\subsection{Close-in free space reference distance}
The Close-in free space reference distance path loss is given by~\ref{eq1} \cite{rappaport2014millimeter,maccartney2014omnidirectional,rappaport2015wideband} 
\begin{align}
	\label{eq1}
	PL(d)[\text{dB}] =& PL(d_{0}) + 10 n \cdot \log_{10} ( \frac{d}{d_{0}} )
	\nonumber
	\\
	&+ \chi_{\sigma}, \textrm{for} \quad d \geq d_{0}
\end{align}
where, $d_{0} $ is the close-in free space reference distance, $\chi_{\sigma}$ is a normal random variable with mean $0$ dB and standard deviation $\sigma$ \cite{rappaport2014millimeter}, $n$ is the PLE, and $ PL(d_{0})$ is the close-in free space path loss in dB that is given by.
\begin{equation}
	\label{eq2}
	PL(d_{0}) = 20\log_{10} \frac{4\pi d_{0}}{\lambda}.
\end{equation}
The parameters of this model are obtained by finding the best minimum mean-square error line fit to the measurement data. In this paper, $d_{0} = 1$ m is used for simplicity.

\subsection{Floating-intercept path loss model}
The floating-intercept (FIM) model is used in the WINNER II and 3GPP~\cite{GPP,kyosti2007winner} channel models, presented in~\eqref{PL_FREQ}.
\begin{align}
	\label{PL_FREQ}
	PL(\text{dB})= \alpha + 10 \beta \log_{10}(d)+ \chi_{\sigma}
\end{align}
where, $\alpha$ is the floating intercept in dB, and $\beta$ is the linear slope, and $\chi_{\sigma}$ is a normal random variable with standard deviation $\sigma$. The LS regression approach \cite{rappaport2014millimeter,mahfuza2017,maccartney2013path} creates a line-of-best fit to the empirical data.

\begin{figure}[t]
	\begin{center}
		
		\includegraphics[width=0.9\linewidth,keepaspectratio]{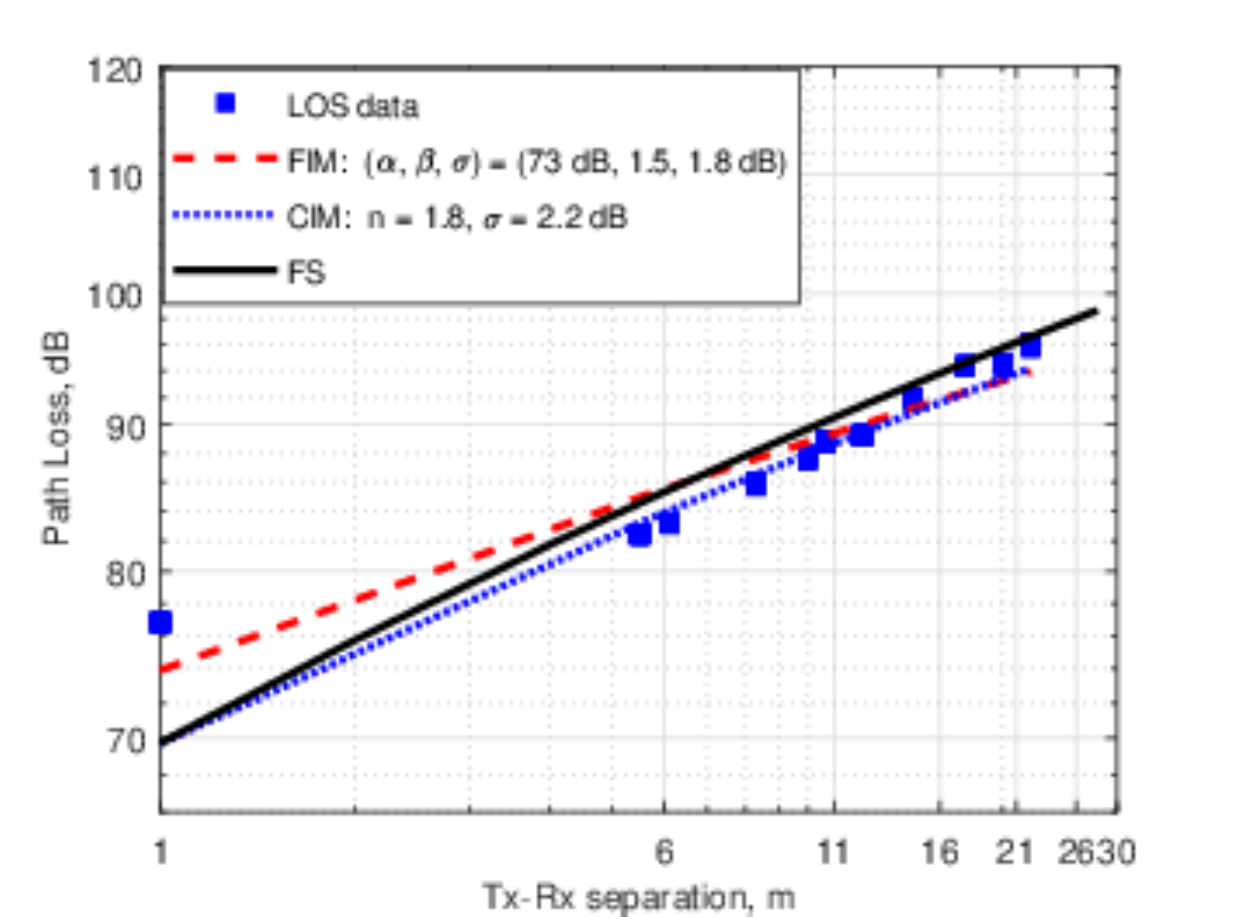}
	\end{center}
	\caption{FIM and CIM along with the measurement data taken from the MEC building in LOS outdoor scenario ($h_{tx} = 1.21$ m and  $h_{rx}= 1.18$ m ).  The blue square and solid black line represent the measurement LOS data and free-space path loss at 73 GHz respectively. The dashed red line and dotted blue show the FIM and CIM path loss models respectively }
	\label{Fig9}
\end{figure}
\section{Propagation Path Loss Results}
\label{sec:results}
The empirical data and the corresponding regression line plots are shown in Figs.~\ref{Fig5} and~\ref{Fig6} for the university building and the airport gate area respectively in the LOS indoor scenario. The PLE (< 2)  may be explained by the presence of a waveguide effect caused by the corridor walls. The shadowing factor is higher in the airport gated area than in the hallway corridor. This can be attributed to the many objects and metallic chairs in the gated area.
The larger path loss values are found at some locations in Fig.~\ref{Fig6}. This is most likely due to the movements of the passengers at the airport gate during taking the measurement. The path loss parameters extracted from FIM and CIM models are tabulated in Table~\ref{tab2}. Here, $\text{h}_{\text{Rx}}$ denotes the height of the receiver antenna, $\text{h}_{\text{Tx}}$ denotes the height of the transmitter antennas. Our results show that the CIM model generates the PLE of 1.8, 2.1, and 1.8 in the building hallway, indoor airport gate, and the outdoor campus areas, respectively which are very close to the free space PLE, $n = 2 $.

Figs.~\ref{Fig7} and~\ref{Fig8} show the path loss values for the LOS indoor and outdoor scenario in the airport gate areas respectively.  The measurement data were captured when the $\text{h}_{\text{Rx}}$ was varied at three heights: $ 1.6$, $1.47$ and $1.3$m from the ground level keeping the $\text{h}_{\text{Tx}}$ at $1.6$ m. The results show that when the Rx antenna height decreases, the slope is decreased, but the shadowing factor increases. The slope values, $\beta$, are found to be $ 0.4$ and $0.62$ for airport indoor and outdoor environments, respectively when the transmit and receive antennas have the smallest height difference. The shadow factors determined for the path loss models developed from the
indoor and outdoor airport gate measurements using the
floating intercept are shown in Table~\ref{tab3}. The results show that the channel has high losses when both antennas are not at the same height.  In addition to the airport outdoor campaign, Fig.~\ref{Fig9} shows the scattered measurement data along with the FIM and CIM for the outdoor campus environment. The CIM provides PLE and shadow factor 1.8 and 2.2 dB respectively in this scenario.

\section{Conclusion}
\label{sec:conclusion}
This paper presented the results of our extensive channel measurement campaign at 73 GHz at Boise Airport, and Boise State University. The FIM and CIM path loss models were developed for both indoor and outdoor airport scenarios. Our work shows that the PLEs from the CIM model are close to the PLE (=2) of the free-space model, whereas the FIM gives a better fit to the measured data. These results also show that the indoor airport environment is uniquely different from other indoor settings due to its large and open nature. Future work includes more measurements at another E-band frequency (81 GHz) for channel parameter estimation.

\section*{Acknowledgment}
This work was funded by NASA$'$s Aeronautics Research Mission Directorate. The authors would like to thank the Boise Airport Operation Team.

\bibliographystyle{IEEEtran}
\bibliography{IEEEabrv,ref}

\end{document}